\documentclass[%
reprint,
superscriptaddress,
nofootinbib,
amsmath,
amssymb,
aps,
prd,
floatfix,
showkeys,
]{revtex4-2}

\usepackage{graphicx,amsfonts,amssymb,amsbsy,amsmath}
\usepackage{slashed}
\usepackage{pstricks}
\usepackage[colorlinks]{hyperref}
\hypersetup{linkcolor=red,citecolor=blue,urlcolor=blue}

\begin{document}

\title{Surface and curvature tensions of cold dense quark matter: \\ a term-by-term analysis within the Nambu-Jona-Lasinio model}

\author{A. G. Grunfeld}
\affiliation{CONICET, Godoy Cruz 2290, Buenos Aires, Argentina}
\affiliation{Departamento de F\'\i sica, Comisi\'on Nacional de  Energ\'{\i}a At\'omica, (1429) Buenos Aires, Argentina.}

\author{M. F. Izzo Villafañe}
\affiliation{Departamento de F\'\i sica, Comisi\'on Nacional de  Energ\'{\i}a At\'omica, (1429) Buenos Aires, Argentina.}

\author{G. Lugones}
\email{german.lugones@ufabc.edu.br}
\affiliation{Universidade Federal do ABC, Centro de Ci\^encias Naturais e Humanas, Avenida dos Estados 5001 - Bang\'u, CEP 09210-580, Santo Andr\'e, SP, Brazil.}

\begin{abstract}
In this paper, we conduct a thorough investigation of the surface and curvature tensions, $\sigma$ and $\gamma$, of three-flavor cold quark matter using the Nambu-Jona-Lasinio (NJL) model with vector interactions. Our approach ensures both local and global electric charge neutrality, as well as chemical equilibrium under weak interactions. By employing the multiple reflection expansion formalism to account for finite size effects, we explore the impact of specific input parameters, particularly the vector coupling constant ratio $\eta_V$, the radius $R$ of quark matter droplets, as well as charge-per-baryon ratio $\xi$ of the finite size configurations. We focus on the role of the contributions of each term of the NJL Lagrangian to the surface and curvature tensions in the mean field approximation. 
We find that the total surface tension exhibits two different density regimes: it remains roughly constant at around $100 \, \mathrm{MeV \, fm^{-2}}$ up to approximately $2-4$ times the nuclear saturation density, and beyond this point, it becomes a steeply increasing function of $n_B$. The total surface and curvature tensions are relatively insensitive to variations in $R$ but are affected by changes in $\xi$ and $\eta_V$. We observe that the largest contribution to $\sigma$ and $\gamma$ comes from the regularized divergent term, making these quantities significantly higher than those obtained within the MIT bag model.
\end{abstract}


\maketitle

\section{Introduction}

One of the major unresolved issues in modern astrophysics is understanding the internal structure and composition of neutron stars. Current theoretical and observational constraints can be explained by significantly different models, suggesting that the core may be composed of nucleonic matter, the appearance of hyperons, quark matter, or even dark matter.

The possibility of quark matter existing in the core of neutron stars is quite compelling, as current understanding of the Quantum Chromodynamics (QCD) phase diagram suggests that, under high density and temperature conditions, nuclear matter would deconfine, releasing its constituent quarks. Indeed, there is some evidence that a quark-gluon plasma has formed in heavy-ion collisions, supporting the idea that deconfinement could also occur under the high-density conditions inside neutron stars. If this is the case, many neutron stars, especially those with higher masses and consequently higher central densities, could be hybrid stars, consisting of a core of quark matter surrounded by hadronic matter.

Although research on hybrid stars has advanced our understanding of their general properties over the past few decades, several critical unresolved issues persist. Among these are the structure and microphysical properties of the region separating the quark and hadron phases.  Studies indicate that,  if finite-size effects, such as surface tension, curvature tension, and Coulomb energy, are sufficiently small, the minimization of the system's energy tends to favor global electrical neutrality over local neutrality \cite{Heiselberg:1999mq, Voskresensky:2002csa, Maruyama:2007hqm, Xia:2019pnq, Yasutake:2019csp, Xia:2020brt}. This results in the formation of what is known as a mixed phase, consisting of electrically charged geometric structures within a background of the other phase with opposite electric charge \cite{Ravenhall:1983uh, Glendenning:1992vb, Glendenning:2001pe, Endo:2006cse}. This mixed phase, often referred to as the quark-hadron \textit{pasta} phase, features structures reminiscent of various types of Italian pasta, such as gnocchi, spaghetti, and lasagna, embedded in a uniform ``sauce" of the other phase that extend over a wide range of densities within the star. On the other hand, if finite-size effects have a high energy cost, a sharp boundary would separate the inner core of quark matter from the outer layers of hadronic matter.

Despite the critical role of surface and curvature tensions, their precise values in the dense environment of neutron stars remain uncertain, as it is not possible to determine them from first-principle calculations in the low-temperature, high-density regime.  Consequently, while various calculations of these parameters exist in the literature, they currently only offer a general evaluation of their potential range, and many of them disagree with each other in their qualitative implications. For example,  using the thin-wall formalism, it has been found that surface tension values are generally low ($\sigma < 30~\mathrm{MeV}/\mathrm{fm}^2$) across various equations of state (EoS) models, such as the NJL model \cite{Garcia:2013eaa, Ke:2013wga, Pinto:2012aq}, the linear sigma model \cite{Palhares:2010be, Pinto:2012aq, Kroff:2014qxa}, and the Polyakov-quark-meson model \cite{Mintz:2012mz}. Similarly, the multiple reflection expansion (MRE) formalism also results in low surface tension values when quark matter is approximated as a free particle gas, which could be a reasonable approximation at extremely high densities due to the property of asymptotic freedom \cite{Berger:1986ps, Lugones:2016ytl, Lugones:2018qgu, Lugones:2020qll}. However, when various interaction channels are taken into account, which better represent the non-perturbative density regime of neutron star matter, the MRE formalism yields significantly higher $\sigma$ values, as seen in studies using the NJL model \cite{Lugones:2013ema} and the vector MIT bag model \cite{Lugones:2021tee}. In contrast, the study of curvature tension is less extensive, although it has been shown that it alters the density range over which each type of geometric structure prevails \cite{Mariani:2023kdu}.

In this context, it is clear that understanding the role of strong interactions in surface and curvature tensions is crucial to fully characterize the behavior and properties of the mixed phase. Although previous works have partially addressed this issue, a comprehensive analysis of how each term in the effective NJL Lagrangian impacts $\sigma$ and $\gamma$, as undertaken in this study, has not yet been performed.
Among the terms we will analyze, the one representing repulsive vector interactions is particularly relevant. Indeed, their impact on $\sigma$ and $\gamma$ has not yet been thoroughly evaluated in the context of the NJL model. These terms are important not only for their ability to stiffen the EOS, thereby explaining the existence of recently observed two-solar-mass pulsars, but also because they can significantly increase surface and curvature tensions, as recently demonstrated in the context of the MIT bag model \cite{Lugones:2021tee}.
To address finite size effects, we will employ the MRE formalism, focusing on matter that is in chemical equilibrium under weak interactions while taking into account both local and global electric charge neutrality. Given these conditions, we will investigate the influence of the vector coupling constant, the radius of quark matter droplets, and the charge-per-baryon ratio.

This paper is organized as follows: Sec. \ref{sec:2} discusses the NJL model in the bulk, including two types of vector interactions. Sec. \ref{sec:3} addresses finite size effects by adopting the MRE formalism, separating the contribution of each term in the NJL Lagrangian to the surface and curvature tensions. Sec. \ref{sec:4} presents our numerical results, examining the impact on the surface and curvature tensions of varying the vector coupling constant, the radius of quark matter droplets, and the charge-per-baryon ratio. Finally, Sec. \ref{sec:5} concludes the paper by summarizing our findings, discussing their implications, and comparing them with previous works.

\section{The model in the bulk}
\label{sec:2}

We begin with an $SU(3)_\mathrm{f}$  NJL effective model at zero temperature, incorporating vector interactions. The specific form of the Lagrangian under study is the following:
\begin{equation}
\begin{aligned}
\mathcal{L} = & ~ \bar{q}(i \slashed{\partial}   - m) q+\tfrac{1}{2} G_S \sum_{a=0}^8\left[\left(\bar{q} \lambda^a q\right)^2+\left(\bar{q} i \gamma_5 \lambda^a q\right)^2\right]  \\
& -G_D\left[\operatorname{det} \bar{q}\left(1+\gamma_5\right) q+   \mathrm{h.c.}\right] \\
& -\left\{\begin{array}{l}
\frac{1}{2} g_V\left(\bar{q} \gamma^\mu q\right)^2 \\
\frac{1}{2} G_V \sum_{a=0}^8\left[\left(\bar{q} \gamma^\mu \lambda^a q\right)^2+\left(\bar{q} i \gamma^\mu \gamma_5 \lambda^a q\right)^2\right] .
\end{array}\right.
\end{aligned}
\end{equation}
In this equation, $q_i (i=u, d, s)$ denotes the quark fields, which are characterized by three colors and flavors, each associated with a corresponding current quark mass $m_i$. We assume $m_u = m_d$. The term with coefficient $G_S$ represents a $U(3)_L \times U(3)_R$ symmetric four-Fermi interaction, where $\lambda^a$ are the Gell-Mann matrices, with $\lambda^0 = \sqrt{2/3}I$. The interaction proportional to $G_D$, known as the Kobayashi-Maskawa-'t Hooft  interaction, accounts for the breaking of $U(1)_A$ symmetry. Regarding the vector channel, we consider two types of contributions, termed Model 1 and Model 2. In Model 1, the term governed by $g_V (>0)$ provides a flavor-independent repulsive interaction, whereas in Model 2, the term associated with $G_V (>0)$ introduces a flavor-dependent repulsion, as discussed in Ref. \cite{Masuda:2012ed}.

Within the mean-field approximation, the thermodynamic potential per unit volume is expressed as the sum of the following components:
\begin{equation}
\Omega_\mathrm{MFA} =   \Omega_\mathrm{div} + \Omega_\mathrm{free} + \Omega_\mathrm{cond} + \Omega_\mathrm{det} + \Omega_\mathrm{vec},
\label{omegaMFAfull}
\end{equation}
where $\Omega_\mathrm{div}$ stands for the regularized divergent contribution, $\Omega_\mathrm{free}$ has the same functional form as the $T \rightarrow 0$ limit of the free Fermi gas, $\Omega_\mathrm{cond}$ represents the condensate term,  $\Omega_\mathrm{det}$ is associated with  the Kobayashi-Maskawa-'t Hooft interaction, and $\Omega_\mathrm{vec}$ represents the vector term.
The explicit form of each contribution is: 
\begin{eqnarray}
\Omega_\mathrm{div} &=& - 6  \sum_{i=u, d, s} \int^{\Lambda}_{0} \frac{k^2 dk}{2 \pi^2}  \ E_{i},  \label{omegaDIV} \\
\Omega_\mathrm{free} &=&  - 6  \sum_{i=u, d, s} \int_{0}^{\kappa_{i}}  \frac{k^2 dk}{2 \pi^2}           ({\mu}^*_{i} - E_{i}) ,  \label{omegaMED} \\
\Omega_\mathrm{cond} &=&  G_S \Big( \phi_u^2 +\phi_d^2  +\phi_s^2 \Big) , \label{omegaCOND} \\
\Omega_\mathrm{det} &=& -4G_D  \phi_u \phi_d  \phi_s ,  \label{omegaDET} \\
\Omega_{\mathrm{vec},1}  &= & -\tfrac{1}{2} g_{V}\Big(  \sum_{i=u, d, s} n_{i}\Big)^{2}  \quad \text{(Model 1)}, \label{eq:Omega_V1}  \\
\Omega_{\mathrm{vec},2} &=  &  -\tfrac{1}{2} G_{V} \sum_{i=u, d, s} n_{i}^{2}  \qquad \text{(Model 2)}. \label{eq:Omega_V2}
\end{eqnarray}
In Eq. \eqref{omegaDIV}, $\Lambda$ represents an ultraviolet cutoff introduced to regularize the divergent integral, and $E_i = \sqrt{k^2 + M_i^2}$ is the on-shell energy of the quark, self-consistently evaluated for the constituent quark masses $M_{i}$, which are dynamically generated via the NJL interactions governed by the coupling constants $G_S$ and $G_D$ \cite{Buballa:2003qv}:
\begin{equation}
M_i = m_i - 2 G_S \phi_i + 2  G_D \phi_j \phi_k.
\label{eq:mass}
\end{equation}
In this equation, $\phi_{i}$ are the quark condensates associated with each flavor. The value of $\phi_i$ is ascertained through a self-consistent solution of the following equation:
\begin{equation}
\phi_{i} = -6 
\int \frac{k^2 dk}{2 \pi^2} \frac{M_{i}}{E_{i}} [1  - \theta(E_{i} - \mu^*_{i}) ],
\label{eq:condensate_definition}
\end{equation}
simultaneously with Eq. \eqref{eq:mass}.

In the preceding expressions, the effective quark chemical potentials $\mu^{*}_{i}$ account for the influence of vector interactions and are defined as follows:
\begin{align}
\mu^{*}_{i} = & \  \mu_{i} - g_V \sum_{j=u,d,s} n_j   & \qquad \text{(Model 1)},  \label{eq:mu_tilde_1} \\
\mu^{*}_{i} = & \  \mu_{i} - G_V n_{i}    & \qquad \text{(Model 2)} , \label{eq:mu_tilde_2}
\end{align}
where $n_{i}$ denotes the particle number densities, calculated as:
\begin{equation}
n_{i}(\mu^*_{i})=6  \int_0^{\kappa_{i}} \frac{ k^2 dk  }{2 \pi^2}.
\label{eq:number_density}
\end{equation}
The Fermi momentum $\kappa_{i}$, used in Eqs. \eqref{omegaMED} and \eqref{eq:number_density} is defined by: 
\begin{equation}
\kappa_{i} = \theta(\mu^*_{i} - M_{i}) \sqrt{{\mu^*_{i}}^2 - M^2_{i}},
\end{equation}
where $\theta$ is the Heaviside step function. 
To determine $\mu^{*}_i$, one must substitute Eq. \eqref{eq:number_density} into Eqs. \eqref{eq:mu_tilde_1} or \eqref{eq:mu_tilde_2} and then find a self-consistent solution for these equations, simultaneously solving Eq. \eqref{eq:mass}.

The portion of Eq. \eqref{omegaMED} containing $E_i$ could be incorporated into $\Omega_\mathrm{div}$ by altering the limits of integration. However, we opt to define $\Omega_\mathrm{free}$ as specified in Eq. \eqref{omegaMED} because, at finite temperature, this definition ensures that the term has the same functional form of a free Fermi gas.

The complete thermodynamic potential is obtained by including the contributions of electrons, denoted as $\Omega_e$, and a vacuum constant, denoted as $\Omega_\mathrm{vac}$. Electrons are modeled as a non-interacting gas of massless fermions, and their thermodynamic potential per unit volume is given by:
\begin{equation}
\Omega_e(T,\mu_{e})   = - \frac{\mu_e^4}{12 \pi^2}.
\end{equation} 
The vacuum constant $\Omega_\mathrm{vac}$ is introduced to ensure that the pressure becomes zero at both zero temperature and zero chemical potentials. Therefore, the complete thermodynamic potential per unit volume is expressed as:
\begin{equation}
\Omega = \Omega_\mathrm{MFA}  + \Omega_e - \Omega_\mathrm{vac}.     \label{QMP}
\end{equation}
It is worth noting that $\Omega_\mathrm{vac}$ does not exert a direct impact on the values of the surface and curvature tensions, as will be elucidated below.

\section{Finite size effects}
\label{sec:3}

To incorporate finite size effects into the thermodynamic potential, we adopt the  MRE formalism (as detailed in \cite{Lugones:2021tee} and its references), which, for the case of a finite spherical droplet, modifies the density of states as follows:
\begin{equation}
\rho_{i}(k) = 1 + \frac{6\pi^2}{kR} f_{S, i}(k) + \frac{12\pi^2}{(kR)^2} f_{C, i}(k) .
 \end{equation}
Here, the surface contribution to the density of states is:
\begin{equation}
f_{S, i}(k)  = - \frac{1}{8 \pi} \left(1 -\frac{2}{\pi} \arctan \frac{k}{m_{i}} \right), 
\end{equation}
and the curvature contribution is given by:
\begin{equation}
f_{C, i}(k)  =  \frac{1}{12 \pi^2} \left[1 -\frac{3k}{2m_{i}} \left(\frac{\pi}{2} - \arctan \frac{k}{m_{i}} \right)\right] .
\end{equation}

The MRE density of states for massive quarks exhibits a reduction compared to the bulk density, leading to negative values within a specific small momentum range. To address this non-physical behavior, an infrared (IR) cutoff in momentum space is introduced. Consequently, the following replacement is necessary to compute the relevant thermodynamic quantities:
\begin{equation}
\int_0^{\Lambda} {{\cdots}} \frac{k^2 \, dk}{2 \pi^2}  \longrightarrow
\int_{\Lambda_\mathrm{IR}}^\Lambda {{\cdots}} \frac{k^2 \, dk}{2 \pi^2} \rho(k).
\label{MRE}
\end{equation}
The infrared cut-off $\Lambda_\mathrm{IR}$ is the largest solution of the
equation $\rho_i (k)= 0$ with respect to the momentum $k$, and it depends on the flavor and the drop's radius as shown in Table \ref{table:cutoff}.

\begin{table}[tb]
\centering
\begin{tabular}{c c c c c }
\hline \hline
particles & $m \, \mathrm{[MeV]}$   & $R \, \mathrm{[fm]}$   & $\Lambda_{\mathrm{IR}} \, \mathrm{[MeV]}$    \\
\hline  
quarks u, d & 5.5     &   3     & 51.77  \\ 
            & 5.5     &   5     & 32.86 \\   
            & 5.5     &  10    & 18.34 \\  

\hline
quarks s &135.7     &   3    & 95.07 \\
         &135.7     &   5    & 62.15 \\   
         &135.7     &  10    & 33.42 \\ 
\hline \hline 
\end{tabular}
\caption{Infrared cutoff $\Lambda_{\mathrm{IR}}$ for different particle masses $m$ and different values of the drop's radius $R$. In the bulk case ($R = \infty$),  $\Lambda_{\mathrm{IR}}$ is zero.} 
\label{table:cutoff}
\end{table}

Introducing the MRE density of states in Eq. \eqref{omegaMFAfull}, the complete expression for the thermodynamic potential of a finite spherical droplet can be expressed as \citep{ Lugones:2020qll, Lugones:2011xv, Lugones:2018qgu, Lugones:2016ytl}:
\begin{align}
\Omega^\mathrm{MRE} V=-P V+\sigma S+\gamma C + \cdots,
\label{finite}
\end{align}
where $V=\frac{4}{3} \pi R^{3}$, $S=4 \pi R^{2}$, $C=8 \pi R$, and the dots represent higher order terms. The coefficients corresponding to the volume, surface, and curvature in Eq. \eqref{finite} are recognized as the pressure ($P$), surface tension ($\sigma$), and curvature tension ($\gamma$) of the droplet, respectively, and are defined as:
\begin{eqnarray}
P  & \equiv & - \frac{\partial \Omega^\mathrm{MRE}}{ \partial V }
\bigg|_{\mu, S, C} , \\
\sigma &\equiv & \frac{\partial \Omega^\mathrm{MRE}}{ \partial S }
\bigg|_{\mu, V, C}  ,  \\
\gamma &\equiv &  \frac{\partial \Omega^\mathrm{MRE}}{ \partial C } \bigg|_{\mu, V, S}  .
\end{eqnarray}

Furthermore, considering the different terms that constitute the mean-field approximation thermodynamic potential given in Eq. \eqref{omegaMFAfull}, it is possible to obtain the contribution of each one of them to the total $P$, $\sigma$ and $\gamma$: 
\begin{eqnarray}
P_\mathrm{tot} &=& P_\mathrm{vec} + P_\mathrm{div} + P_\mathrm{free} + P_\mathrm{cond} + P_\mathrm{det},   \label{eqq:total_pressure} \\
\sigma_\mathrm{tot} &=& \sigma_\mathrm{vec} + \sigma_\mathrm{div} + \sigma_\mathrm{free} + \sigma_\mathrm{cond} + \sigma_\mathrm{det},   \label{eqq:total_surface} \\
\gamma_\mathrm{tot} &=& \gamma_\mathrm{vec} + \gamma_\mathrm{div} + \gamma_\mathrm{free} + \gamma_\mathrm{cond} + \gamma_\mathrm{det} . \label{eqq:total_curvature}
\end{eqnarray}
In the following subsections, we will deduce each of the previously mentioned contributions in detail.

\subsection{Divergent term}

By taking the divergent term from Eq. \eqref{omegaDIV} and adopting the procedure from Eq. \eqref{MRE}, we obtain:
\begin{equation}
\Omega_{\mathrm{div},i}^{\mathrm{MRE}} =- 6 \int^{\Lambda}_{{\Lambda_{\mathrm{IR},i}}} \frac{k^2 dk}{2 \pi^2} E_{i} \rho_{i} .
\label{omegaDIV_MRE}
\end{equation}
Proceeding as in Eq. \eqref{finite}, that is, by separating the terms proportional to $V$, $S$, and $C$, we identify the pressure, the surface tension, and the curvature tension as follows:
\begin{eqnarray}
P_{\mathrm{div},i} &=& 6 \int^{\Lambda}_{{\Lambda_{\mathrm{IR},i}}}  \frac{k^2 dk}{2 \pi^2} \, E_{i},    \\ 
\sigma_{\mathrm{div},i} &=&- 6 \int^{\Lambda}_{{\Lambda_{\mathrm{IR},i}}} k \, dk  E_{i} f_{S,i},  \\
\gamma_{\mathrm{div},i} &=& - 6 \int^{\Lambda}_{{\Lambda_{\mathrm{IR},i}}} dk E_{i} f_{C,i} .
\end{eqnarray}

\subsection{Free term}

Applying the procedure from Eq. \eqref{MRE} to Eq. \eqref{omegaMED}, we obtain:
\begin{equation}
\Omega_{\mathrm{free},i}^{\mathrm{MRE}} =  - 6   \int_{{\Lambda_{\mathrm{IR},i}}}^{\kappa_{i}}  \frac{k^2 dk}{2 \pi^2} \; ({\mu}^*_{i} - E_{i}) \,\rho_{i}  .
\label{omegaFREE}
\end{equation}
Splitting the previous expression as in Eq. \eqref{finite} we find:
\begin{eqnarray}
P_{\mathrm{free},i} &=& 6   \int^{\kappa_{i}}_{{\Lambda_{\mathrm{IR},i}}}  \frac{k^2 dk}{2 \pi^2}  ({\mu}^*_{i} - E_{i})  , \\
\sigma_{\mathrm{free},i} &=&- 6  \int^{\kappa_{i}}_{{\Lambda_{\mathrm{IR},i}}}   k \,  dk ({\mu}^*_{i} - E_{i}) f_{S,i}  , \\
\gamma_{\mathrm{free},i} &=&  -6  \int^{\kappa_{i}}_{{\Lambda_{\mathrm{IR},i}}} dk \; ({\mu}^*_{i} - E_{i}) \,f_{C,i} .
\end{eqnarray}

\subsection{Condensate term}

The condensate term is given in Eq. \eqref{omegaCOND} where the condenstates $\phi_i$ are defined in Eq. \eqref{eq:condensate_definition}.  Using Eq. \eqref{MRE}, the condensate contribution reads:
\begin{equation}
\Omega_{\mathrm{cond}}^{\mathrm{MRE}} =  G_S \left[ \left( \phi_{u}^{\mathrm{MRE}} \right)^2 + \left(  \phi_{d}^{\mathrm{MRE}} \right)^2 + \left( \phi_{s}^{\mathrm{MRE}} \right)^2 \right] 
\label{omegaCOND_MRE}
\end{equation}
where
\begin{eqnarray}
\phi_{i}^{\mathrm{MRE}} &=& -6 \int^{\Lambda}_{{\Lambda_{\mathrm{IR},i}}} \frac{k^2 dk}{2 \pi^2}   \frac{M_{i}}{E_{i}} [1  - \theta( \mu^*_{i} - E_{i} ) ]  \rho_{i} \\
&=&  -6 \int^{\Lambda}_{{\Lambda_{\mathrm{IR},i}}} \frac{k^2 dk}{2 \pi^2}  \frac{M_{i}}{E_{i}} \rho_{i} + 6 \int^{\kappa_{i}}_{{\Lambda_{\mathrm{IR},i}}} \frac{k^2 dk}{2 \pi^2}  \frac{M_{i}}{E_{i}} \rho_{i}  \qquad \\
&= & - 6 \int_{\kappa_{i}}^{\Lambda} \frac{k^2 dk}{2 \pi^2} \frac{M_{i}}{E_{i}} \rho_{i} .
\label{eq:condensate_with_MRE}
\end{eqnarray}
Proceeding in the same way as in Eq. \eqref{finite}, the condensates take on the following form:
\begin{equation}
\phi_{i}^{\mathrm{MRE}}=\phi_{i}^{V}+\frac{S}{V} \phi_{i}^{S}+\frac{C}{V} \phi_{i}^{C},
\label{eq:phi_MRE}
\end{equation}
where we define:
\begin{eqnarray}
\phi_{i}^{V} & \equiv & -6  \int_{\kappa_{i}}^{\Lambda}  \frac{k^2 dk}{2 \pi^2}  \frac{M_{i}}{E_{i}} , \label{eqq:phi_V}\\ 
\phi_{i}^{S} & \equiv & -6  \int_{\kappa_{i}}^{\Lambda}  k d k  \frac{M_{i}}{E_{i}} f_{S, i}, \label{eqq:phi_S} \\ 
\phi_{i}^{C} & \equiv & -6  \int_{\kappa_{i}}^{\Lambda} dk \frac{M_{i}}{E_{i}} f_{C, i}.
\end{eqnarray}
Replacing Eq. \eqref{eq:phi_MRE} into Eq. \eqref{omegaCOND_MRE} we obtain:
\begin{equation}
\begin{aligned}
\Omega_{\mathrm{cond}}^{\mathrm{MRE}} = & ~  G_S \sum_i \left[ (\phi_i^V)^2 + \frac{S^2}{V^2} (\phi_i^S)^2 + \frac{C^2 }{V^2} (\phi_i^C)^2  \right. \\ 
& + \left. 2 \frac{S}{V}\phi_i^{V}\phi_i^{S} + 
2 \frac{C}{V}\phi_i^{V}\phi_i^{C} + 2 \frac{SC}{V^2}\phi_i^{S}\phi_i^{C} \right]  .
\end{aligned}
\end{equation}

We now multiply the preceding expression by the volume $V$, retaining only the terms that are proportional to $R^3$, $R^2$, and $R$, and disregarding any other powers of the radius $R$. The outcome is as follows:
\begin{eqnarray}
V \Omega_{\mathrm{cond}}^{\mathrm{MRE}} &=&  G_S \sum_i \left\{V (\phi_i^V)^2 + 
S  2  \phi_i^{V} \phi_i^{S} \right. \nonumber \\
&+& \left. C  \left( \tfrac{3}{2} (\phi_i^S)^2 + 2 \phi_i^{V}\phi_i^C \right) + \cdots \right]  .
\end{eqnarray}
From the above equation, we can easily identify the pressure, surface tension, and curvature tension associated with the condensate term, which turn out to be:
\begin{eqnarray}
P_{\mathrm{cond}} &=& - G_S \sum_i (\phi_i^{V})^2 , \\
\sigma_{\mathrm{cond}} &=& 2 G_S \sum_i \phi_i^{V} \phi_i^{S},  \label{eqq:sigma_cond} \\
\gamma_{\mathrm{cond}} &=& G_S \sum_i \left( \tfrac{3}{2} (\phi_i^{S})^2 + 2 \phi_i^{V}\phi_i^{C} \right) . 
\end{eqnarray}

\subsection{Determinant term} 

The contributions of the determinant term can be derived similarly to the preceding subsection. Starting with
\begin{equation}
\Omega_{\mathrm{det}}^{\mathrm{MRE}} = -4 G_D 
\phi_{u}^{\mathrm{MRE}} \phi_{d}^{\mathrm{MRE}} \phi_{s}^{\mathrm{MRE}},
\label{omegaDET_2}
\end{equation}
replacing Eq. \eqref{eq:phi_MRE}, and multiplying by the volume we arrive at:
\begin{equation}
\begin{aligned}
V \Omega_{\mathrm{det}}^{\mathrm{MRE}} =& -4 G_D \times  \big\{ V \left( \phi_u^V \phi_d^V \phi_s^V\right) \\
&+  S \left[ \phi_u^V \phi_d^V \phi_s^S + \phi_u^V \phi_d^S \phi_s^V + \phi_u^S \phi_d^V \phi_s^V\right] \\ 
&+  C \big[ \phi_u^V \phi_d^V \phi_s^C + \phi_u^V \phi_d^C \phi_s^V + \phi_u^C \phi_d^V \phi_s^V   \\
&+  \tfrac{3}{2}(\phi_u^V \phi_d^S \phi_s^S + \phi_u^S \phi_d^S \phi_s^V + \phi_u^S \phi_d^V \phi_s^S) \big]  \\
& + \cdots   \big\} ,
\end{aligned}
\end{equation}
where terms of order $R^0$ were disregarded. From the latter expression we obtain:
\begin{eqnarray}
P_{\mathrm{det}} &= & 4 G_D \; \phi_u^V \phi_d^V \phi_s^V  , \\
\sigma_{\mathrm{det}} & = & -4 G_D  \left[ \phi_u^V \phi_d^V \phi_s^S + \phi_u^V \phi_d^S \phi_s^V + \phi_u^S \phi_d^V \phi_s^V\right] , \quad  \\
\gamma_{\mathrm{det}} & = & -4 G_D  \left[ \phi_u^V \phi_d^V \phi_s^C + \phi_u^V \phi_d^C \phi_s^V + \phi_u^C \phi_d^V \phi_s^V \right.  \nonumber \\
& &+ \left. 
\tfrac{3}{2}(\phi_u^V \phi_d^S \phi_s^S + \phi_u^S \phi_d^S \phi_s^V + \phi_u^S \phi_d^V \phi_s^S)
\right] .
\end{eqnarray}

\subsection{Vector term in Model 1}

To analyze the contributions of the vector term in Model 1, we  begin by:  
\begin{equation}
\begin{aligned}
\Omega_{\mathrm{vec},1}^{\mathrm{MRE}} = &   - \tfrac{1}{2} g_{V}\left(   n_u^{\mathrm{MRE}} + n_d^{\mathrm{MRE}}  + n_s^{\mathrm{MRE}}    \right)^{2} \\
=  &   - \tfrac{1}{2} g_{V}  \bigg[  (n_u^{\mathrm{MRE}})^2 + (n_d^{\mathrm{MRE}})^2  + (n_s^{\mathrm{MRE}})^2  \\  
 & \qquad \quad +  2 n_u^{\mathrm{MRE}} n_d^{\mathrm{MRE}}  +  2 n_u^{\mathrm{MRE}} n_s^{\mathrm{MRE}} \\
 & \qquad \quad + 2 n_d^{\mathrm{MRE}} n_s^{\mathrm{MRE}}\bigg] .
\end{aligned}
\label{eq:Omega_V1MRE}
\end{equation}
In the above expression,  the particle number density in the MRE formalism is given by:
\begin{align}
n_{i}^{\mathrm{MRE}}=  6  \int^{\kappa_{i}}_{{\Lambda_{\mathrm{IR},i}}}  \frac{ k^2 dk}{2 \pi^2}  \rho_{i}  .
\end{align}
After substituting the MRE density of states $\rho_{i}$ and separating the contributions from volume, surface, and curvature, the particle number density reads:
\begin{align}
n_{i}^{\mathrm{MRE}}=n_{i}^{V}+\frac{S}{V} n_{i}^{S}+\frac{C}{V} n_{i}^{C} ,
\label{eq:n_MRE}
\end{align}
where we define:
\begin{align}
n_{i}^{V} & \equiv 6 \int^{\kappa_{i}}_{{\Lambda_{\mathrm{IR},i}}}  \frac{ k^2 dk}{2 \pi^2}    \label{eq:n_MRE_V} , \\ 
n_{i}^{S} & \equiv 6 \int^{\kappa_{i}}_{{\Lambda_{\mathrm{IR},i}}}   k d k    f_{S, i}  ,  \label{eq:n_MRE_S}\\ 
n_{i}^{C} & \equiv 6 \int^{\kappa_{i}}_{{\Lambda_{\mathrm{IR},i}}}   d k  f_{C, i}    \label{eq:n_MRE_C} .
\end{align}
Replacing Eq. \eqref{eq:n_MRE} into Eq. \eqref{eq:Omega_V1MRE} we find:    
\begin{equation}
\begin{aligned}
\Omega_{\mathrm{vec},1}^{\mathrm{MRE}}  V = -  P_{\mathrm{vec},1} V + \sigma_{\mathrm{vec},1} S +  \gamma_{\mathrm{vec},1} C  + \cdots 
\end{aligned}
\end{equation}
being
\begin{eqnarray}
P_{\mathrm{vec},1} &= & \tfrac{1}{2} g_V \big[ (n^V_u)^2 + (n^V_d)^2  + (n^V_s)^2  \nonumber \\ 
&& +  2 n^V_u n^V_d  + 2 n^V_u n^V_s + 2 n^V_d n^V_s   \big] \nonumber \\
& = &   \tfrac{1}{2} g_V   \bigg( \sum_{i= u,  d, s}    n^V_i    \bigg)^2 ,\\
\sigma_{\mathrm{vec},1} & =&    - g_V   \sum_{i, j = u,  d, s}  n_{i}^{V} n_{j}^{S}  , \\
\gamma_{\mathrm{vec},1}  & = &- g_{V} \sum_{u,d,s} n_{i}^{V} n_{j}^{C} 
  -\tfrac{3}{4} g_{V} \sum_{u, d, s} n_{i}^{S} n_{j}^{S}  .
\end{eqnarray}

\subsection{Vector term in Model 2}

The vector term in Model 2 is given by:  
\begin{equation}
\begin{aligned}
\Omega_{\mathrm{vec},2}^{\mathrm{MRE}} =   - \tfrac{1}{2} G_{V}  \sum_{i=u,d,s}\left(   n_i^{\mathrm{MRE}} \right)^{2}  .
\end{aligned}
\label{eq:Omega_V2MRE}
\end{equation}
Replacing the particle number density $n_i^{\mathrm{MRE}}$ from Eq. \eqref{eq:n_MRE} we obtain:
\begin{equation}
\begin{aligned}
\Omega_{\mathrm{vec},2}^{\mathrm{MRE}}  V = -  P_{\mathrm{vec},2} V + \sigma_{\mathrm{vec},2} S +  \gamma_{\mathrm{vec},2} C  + \cdots 
\end{aligned}
\end{equation}
being
\begin{eqnarray}
P_{\mathrm{vec},2} &=&  \tfrac{1}{2} G_V    \sum_{i= u,  d, s}    \left( n^V_i    \right)^2 ,\\
\sigma_{\mathrm{vec},2}  &=&    - G_V   \sum_{i = u,  d, s}  n_{i}^{V} n_{i}^{S}, \\
\gamma_{\mathrm{vec},2} & = & - G_{V} \sum_{i= u, d, s} n_{i}^{V} n_{i}^{C}  -   \tfrac{3}{4} G_V  \sum_{i = u, d, s} \left( n_{i}^{S}  \right)^2 . \quad
\end{eqnarray}
%

\subsection{Global/local charge neutrality and chemical equilibrium}

In this work, we are interested in studying finite size droplets in $\beta$-equilibrium that may form, for example, within the mixed phase of a hybrid star. In such cases, chemical equilibrium is maintained through weak interactions among quarks, such as $d \rightarrow u + e^- + \bar{\nu}_e$, $u + e^-   \rightarrow  d + \nu_e$, $s \rightarrow u + e^- + \bar{\nu}_e$, and $u + d \leftrightarrow u + s$. Since we are considering a system at zero temperature, neutrinos  leave the system freely, resulting in a vanishing chemical potential  for the neutrinos. Therefore, the chemical equilibrium conditions read:
\begin{eqnarray}
\mu_{d} = \mu_{s} = \mu_{u} + \mu_e,  
\label{beta}
\end{eqnarray}
which can be expressed as:
\begin{eqnarray}
\mu_u &=&  \mu_q - \tfrac{2}{3} \mu_e , \label{eq:chemical_equilibrium_01}\\
\mu_d &=& \mu_q + \tfrac{1}{3}\mu_e , \label{eq:chemical_equilibrium_02} \\
\mu_s &=& \mu_q + \tfrac{1}{3}\mu_e \label{eq:chemical_equilibrium_03},
\end{eqnarray}
where $\mu_q$ is the quark chemical potential. 
The electric charge per unit volume is:
\begin{equation}
n_{Q}=\tfrac{2}{3} n^{\mathrm{MRE}}_{u}-\tfrac{1}{3} n^{\mathrm{MRE}}_{d}-\tfrac{1}{3} n^{\mathrm{MRE}}_{s}-n_{e} .
\label{eq:charge_per_unit_volume}
\end{equation}
For convenience, we will write the charge density in terms of the charge-per-baryon ratio:
\begin{equation}
\xi \equiv \frac{n_{Q}}{n_{B}} ,
\end{equation}
where 
\begin{equation}
n_{B} = \tfrac{1}{3} (n_u + n_d + n_s)
\end{equation}
is the baryon number density. Replacing $n_Q = \xi (n_u+ n_d+n_s)/3$ in Eq. \eqref{eq:charge_per_unit_volume} we obtain the electric charge conservation equation:
\begin{equation}
0=\left( \tfrac{2}{3} -\xi\right) n^{\mathrm{MRE}}_{u}-\left( \tfrac{1}{3} + \xi\right) n^{\mathrm{MRE}}_{d}-\left( \tfrac{1}{3} + \xi\right) n^{\mathrm{MRE}}_{s}-n_{e}.
\label{eq:global_and_local_charge neutrality}
\end{equation}
The calculation of the surface and curvature tensions presumes that there is a boundary separating the quark from the hadronic phase. Across this boundary, the electron background is uniform, i.e., it is the same at the internal and the external side of the drop's boundary, simply because electrons do not feel the strong interaction. Additionally, the hadronic phase is predominantly positive since it is constituted mainly of neutrons and protons. Therefore, due to global charge neutrality, the quark phase inside the drop has to be negative. Thus, $n_Q$ must be negative at the inner side of the drop, i.e. $\xi \leq 0$ (cf. Fig. 5 of Ref. \cite{Mariani:2023kdu}). From Ref. \cite{Mariani:2023kdu} we learn that typical values of $\xi$ are in the range $-0.75 < \xi \leq  0$ (see also Ref. \cite{Wu:2017xaz}).

\section{Numerical results}
\label{sec:4}

In this study, we have adopted the HK parameter set as described in Ref. \cite{Masuda:2012ed}, with the following specific values: $m_{u,d} = 5.5 \, \mathrm{MeV}$, $m_s = 135.7  \, \mathrm{MeV}$, $\Lambda = 631.4  \, \mathrm{MeV}$, $G_S\Lambda^2$ = 3.67, and $G_D \Lambda^5$ = 9.29. Using these parameters, we have computed the corresponding IR cutoff for each flavor and size, as detailed in Table \ref{table:cutoff}. In the figures we will use the parameter $\eta_V$ which represents both $\eta_V = g_V/G_S$ (Model 1) and $\eta_V = G_V/G_S$ (Model 2).  

The thermodynamic potential is determined for each $\mu_q$ as a function of the condensates and the chemical potentials. From this, the thermodynamic quantities of interest can be calculated. Therefore, we numerically solve the set of Eqs. \eqref{eq:mass} self-consistently, along with Eqs. \eqref{eq:mu_tilde_1} or \eqref{eq:mu_tilde_2}, supplemented with the global/local electric charge neutrality condition (Eq. \eqref{eq:global_and_local_charge neutrality}) and the chemical equilibrium conditions (Eqs. \eqref{eq:chemical_equilibrium_01}-\eqref{eq:chemical_equilibrium_03}).
It is important to remark that, in general, when we numerically solve the set of equations related with all the conditions discussed above, there might be regions for which there is more than one solution for each value of  $\mu_q$. To choose the stable solution among all of them, we require it to be an overall minimum of the thermodynamic potential.

\subsection{Masses and densities}

\begin{figure}[tb]
\centering
\includegraphics[scale=0.55, angle=-90]{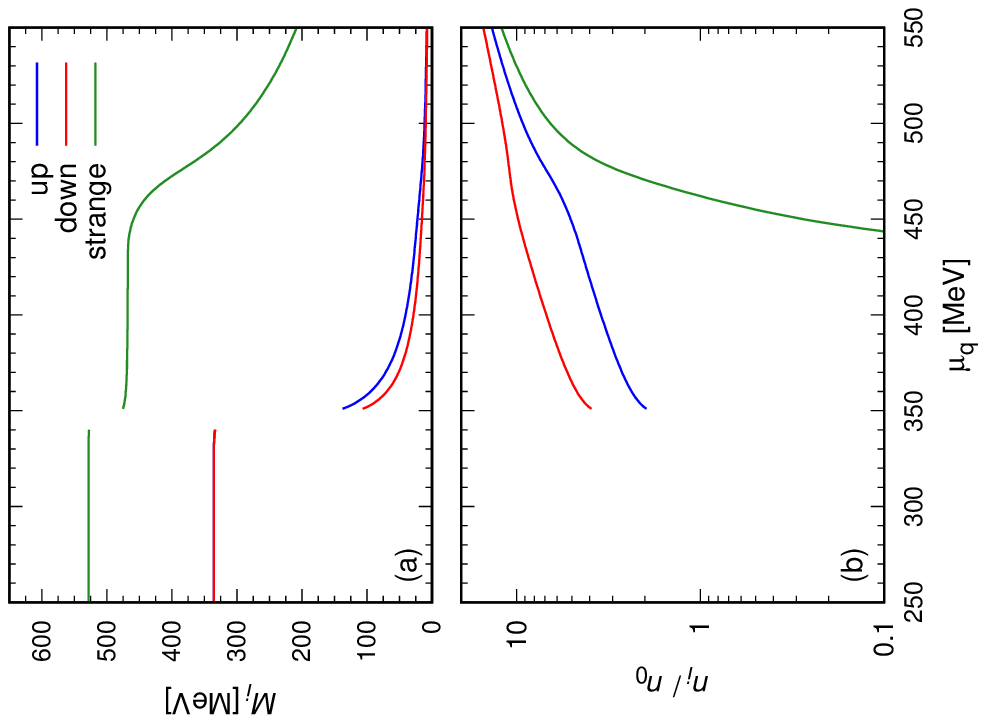}
\caption{Constituent quark masses and quark number densities as functions of the quark chemical potential $\mu_q$, for the bulk case ($R=\infty$), maintaining local electric charge neutrality ($\xi=0$) and excluding vector interactions ($\eta_V=0$). As vector interactions are omitted, there is no distinction between Model 1 and Model 2. The quark number densities are expressed in units of the nuclear saturation density, $n_0 = 0.15 \mathrm{~fm^{-3}}$.}
\label{fig:constituent_masses}
\end{figure}

\begin{figure*}[tbh]
\centering
\includegraphics[scale=1.2]{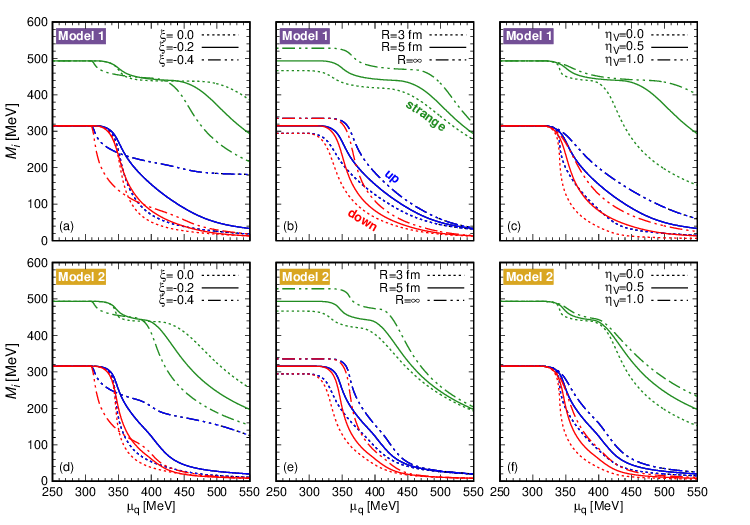}
\caption{Impact of the parameters $\xi$, $R$, and $\eta_V$ on the behavior of the constituent masses $M_{i}$ for each flavor, for the two adopted models of vector interactions (labeled as Model 1 and Model 2). In (a) and (d), the variation of $M_{i}$ with $\mu_q$ is shown for different values of $\xi$, keeping $R = 5 \, \mathrm{fm}$ and $\eta_V = 0.5$. In (b) and (e), changes across different $R$ are examined, keeping $\xi = -0.2$ and $\eta_V = 0.5$. In (c) and (f), the influence of $\eta_V$ is explored, while keeping $\xi = -0.2$ and $R = 5 \, \mathrm{fm}$. The solid curves represent the reference models characterized by $R = 5 \, \mathrm{fm}$, $\xi = -0.2$, and $\eta_V = 0.5$. Therefore, these curves are repeated across (a), (b), and (c) for Model 1, and (d), (e), and (f) for Model 2.}
\label{fig:masses}
\end{figure*}

\begin{figure*}[tbh]
\centering
\includegraphics[scale=1.2]{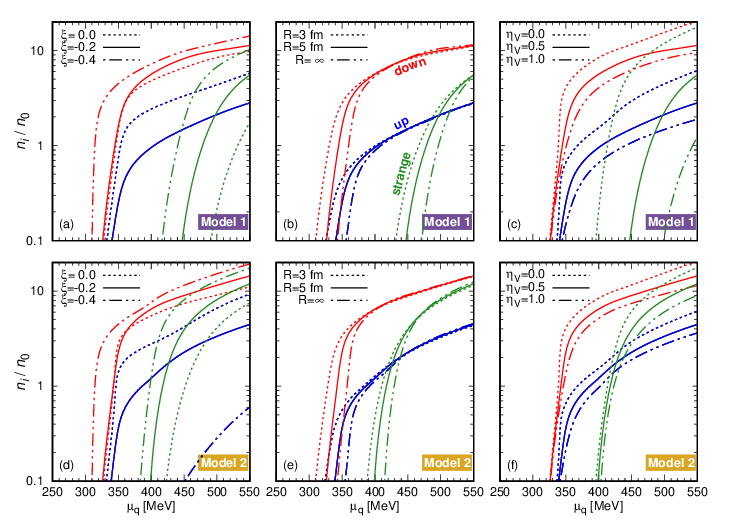}
\caption{Number density for each quark flavor as a function of  $\mu_q$, for different values of the parameters $\xi$, $R$, and $\eta_V$. The figure follows the same organization as the previous one, with the same models and parameters being explored.  In (a) and (d), $\xi$ varies with $R = 5 \, \mathrm{fm}$ and $\eta_V = 0.5$ constant; in (b) and (e), different radii are explored with $\xi = -0.2$ and $\eta_V = 0.5$; and in (c) and (f), different $\eta_V$ values are used with $\xi = -0.2$ and $R = 5 \, \mathrm{fm}$. Solid curves represent reference models with $R = 5 \, \mathrm{fm}$, $\xi = -0.2$, and $\eta_V = 0.5$.}
\label{fig:densities}
\end{figure*}

In Fig. \ref{fig:constituent_masses} we begin by presenting results that are already well-known in the literature \cite{Buballa:2003qv} for the bulk case ($R=\infty$), with local electrical charge neutrality ($\xi=0$) and without vector interactions ($\eta_V=0$). These results will be subsequently used for comparison when we consider different inputs for the parameters $R$, $\xi$, and $\eta_V$.  At $\mu_q = \mu_c \sim 350 \mathrm{~MeV}$, we observe a first-order phase transition, where $M_u$ and $M_d$ decrease sharply from $335 \mathrm{~MeV}$ to a significantly lower value (see Fig. \ref{fig:constituent_masses}(a)). At this transition, the particle number density of $u$ and $d$ quarks jumps from zero to a finite value of approximately $2-3 n_0$, while the particle number density of $s$ quarks remains zero (see Fig. \ref{fig:constituent_masses}(b)). However, due to flavor mixing, $M_s$ does not remain constant but instead exhibits a small jump at $\mu_q = \mu_c$. Beyond $\mu_{\mathrm{c}}$, the contributions of $\phi_u$ and $\phi_d$ to $M_s$ gradually shrink, and $M_s$ remains nearly constant until $\mu_q$ surpasses $M_s$, leading to a non-zero $n_s$ as well. At that point, we observe a smooth crossover above the strange quark threshold.

In Fig. \ref{fig:masses}, we present the constituent quark masses $M_{i}$ as a function of the quark chemical potential $\mu_q$ for quark drops of finite radii ($R = 3, 5 \mathrm{~fm}$) and the bulk scenario ($R = \infty$). We examine conditions with local electric charge neutrality ($\xi=0$) and global charge neutrality (considering $\xi = -0.2, -0.4$). Additionally, the influence of vector interactions is explored with $\eta_V$ values of $0$, $0.5$, and $1$. These analyses are conducted for both Model 1 and Model 2, encompassing all three quark flavors. 
Up, down, and strange quarks are represented by blue, red, and green curves respectively. Each panel illustrates the effect of varying a single parameter while holding the others constant. The curve characterized by $\xi = -0.2$, $R = 5 \, \mathrm{fm}$, and $\eta_V = 0.5$ is depicted with a solid line and is consistently repeated across all the panels. As in the baseline scenario depicted in Fig. \ref{fig:constituent_masses}(a), all curves asymptotically converge toward the current mass of the quarks.
The up and down quarks undergo a continuous partial restoration of chiral symmetry, in contrast to the baseline scenario of Fig. \ref{fig:constituent_masses}(a), which shows a first-order transition.  On the other hand, the strange quark consistently restores chiral symmetry through a crossover mechanism, akin to the pattern observed in the baseline case.

In Figs. \ref{fig:masses}(a) and \ref{fig:masses}(d) it is shown how the curves $M_{i}$ versus $\mu_q$ are affected by changes in the parameter $\xi$ keeping $R = 5 \, \mathrm{fm}$  and $\eta_V = 0.5$. As $\xi$ is changed from zero to increasingly negative values, there is a shift in the behavior of the constituent masses across both models shown.
In the chiral symmetry broken phase, at low chemical potentials, the constituent masses $M_{i}$ align with the baseline model across all $\xi$ values (up to $\mu_q \sim 300\, \mathrm{MeV}$). As $\mu_q$ increases, there is a common qualitative behavior for the $d$ and $s$ quark masses, distinct from that of the $u$ quark mass.
For the $d$ and $s$ flavors, as $\xi$ becomes more negative, the constituent masses decrease for a fixed $\mu_q$. Indeed, chiral symmetry restoration occurs earlier for more negative $\xi$ values. However, the behavior for the $u$ flavor is the opposite. At a fixed chemical potential, the constituent mass of the $u$ quark increases as $\xi$ becomes more negative. Notably, in the most extreme case of $\xi$, chiral symmetry partial restoration for the $u$ quark occurs at very high chemical potentials, after the restoration of $d$ and $s$ quarks. This effect is pronounced in Model 1. As we will discuss later, this behavior is related to the electric charge of each flavor. By imposing a more negative $\xi$, we require a suppression of $u$ quarks, which results in an increase in their mass.

In Figs. \ref{fig:masses}(b) and \ref{fig:masses}(e), we show the behavior of the constituent masses $M_{i}$ while varying $R$, keeping $\xi = -0.2$ and $\eta_V = 0.5$ constant. We observe that as $R$ decreases, the effective mass exhibits a reduction. This effect is more pronounced at lower chemical potentials and diminishes asymptotically. Thus, we see that chiral symmetry partial restoration for all flavors is slightly anticipated when the system size decreases. 

In Figs. \ref{fig:masses}(c) and \ref{fig:masses}(f), we present the behavior of the constituent masses $M_{i}$ while varying $\eta_V$, with $\xi = -0.2$ and $R = 5 \, \mathrm{fm}$ held constant. We observe that as $\eta_V$ increases, the effective mass shows a slight increase. This effect is more pronounced at intermediate chemical potentials and diminishes asymptotically. As is well known, increasing the vector interaction delays the partial restoration of chiral symmetry.

In Fig. \ref{fig:densities}, we show the quark number densities $n_{i}$ as a function of the quark chemical potential $\mu_q$ for the same parameter choices of  Fig. \ref{fig:masses}. As before, each panel illustrates the effect of varying a single parameter while holding the others constant. Similarly, the curve characterized by $\xi = -0.2$, $R = 5 \, \mathrm{fm}$, and $\eta_V = 0.5$ is depicted with a solid line and is repeated across all the panels. As in the baseline scenario, the particle number densities are zero at sufficiently low $\mu_q$; however, they increase gradually without the discontinuity seen in Fig. \ref{fig:constituent_masses}(b).

\begin{figure*}[tbh]
\centering
\includegraphics[scale=1.3]{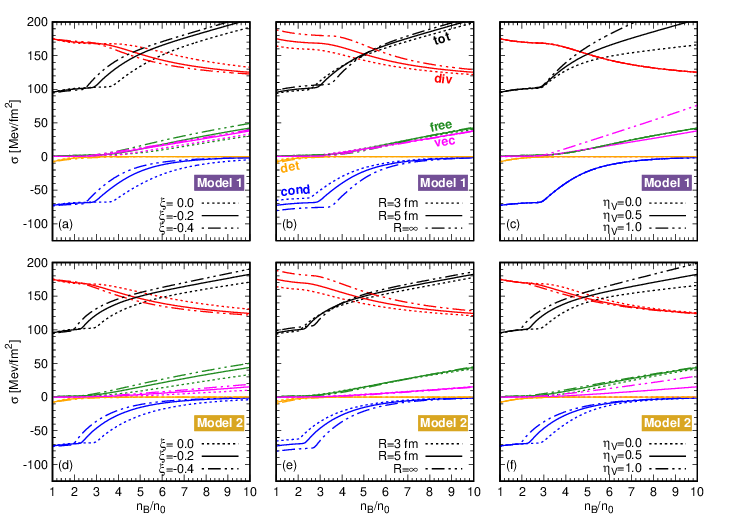}
\caption{Total surface tension (black curves) and its contributions from divergent (red), free (green), condensate (blue), determinant (orange), and vector (magenta) terms as functions of the baryon number density. The parameters $\xi$, $R$, and $\eta_V$ are varied in the same manner as in the previous figure.}
\label{fig:surface_tension}
\end{figure*}

Figs. \ref{fig:densities}(a) and \ref{fig:densities}(d) illustrate how the $n_{i}$ versus $\mu_q$ curves are influenced by variations in the parameter $\xi$, while maintaining $R = 5 \, \mathrm{fm}$ and $\eta_V = 0.5$. For $\xi=0, -0.2$, the onset of $u$ and $d$ quarks occurs at roughly the same value of $\mu_q$, while $s$ quarks appear at larger $\mu_q$. For the more extreme case $\xi = -0.4$, $u$ quarks appear at a much higher chemical potential, even larger than the $s$ quark onset.
For a fixed chemical potential, as the value of $\xi$ becomes more negative, the amount of $d$ and $s$ quarks increases. Comparatively, there are always more $d$ quarks than $s$ quarks simply because the former has a lower mass. The behavior of $u$ quarks is the opposite. As $\xi$ becomes more negative, the amount of $u$ quarks decreases. This behavior is related to the electric charge of each flavor. Imposing a more negative $\xi$ disfavors the $u$ quarks, which are the only ones with a positive charge.
In Figs. \ref{fig:densities}(b) and \ref{fig:densities}(e) we show the role of the drop's radius on the particle number densities. While changes in $R$ affect the $n_i$, these effects are significant only around the onset for each flavor. As $\mu_q$ increases, all curves tend to the bulk case.
Finally, Figs. \ref{fig:densities}(c) and \ref{fig:densities}(f) illustrate how the curves change with variations in the parameter $\eta_V$.  Changes in $\eta_V$ affect the $n_i$, but these effects are not significant around the onset for each flavor. However, as $\mu_q$ increases the effect of vector repulsive interactions becomes substantial and, for a fixed $\mu_q$, reduces the particle number densities as $\eta_V$ increases. This effect is more pronounced for Model 1.

\begin{figure*}[tb]
\centering
\includegraphics[scale=1.3]{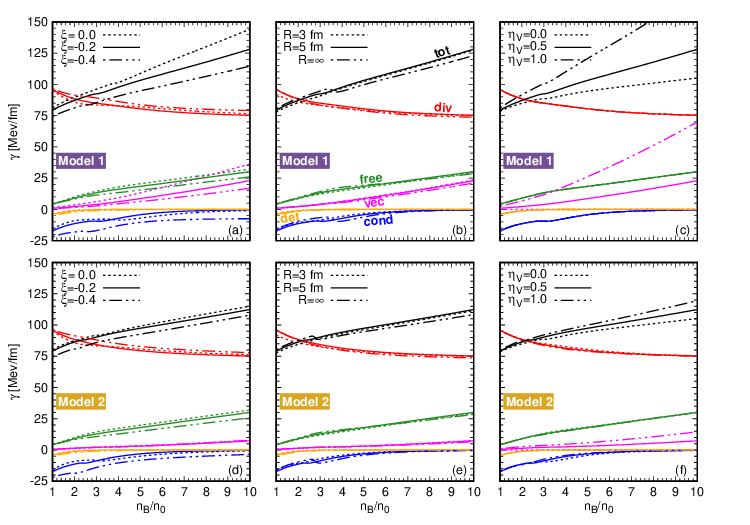}
\caption{Curvature tension as a function of the baryon number density for the same models and parameters of the previous figures.}
\label{fig:curvature_tension}
\end{figure*}

\subsection{Surface and curvature tensions}

In Figs. \ref{fig:surface_tension} and \ref{fig:curvature_tension}, we illustrate the behavior of the surface and curvature tensions as functions of the baryon number density $n_B$ for the same parametrizations of previous figures.

Within each panel of Fig. \ref{fig:surface_tension}, the total surface tension (black curves) is shown along with the contributions from the divergent, free, condensate, determinant, and vector terms, as specified in Eq. \eqref{eqq:total_surface}.    
Note that $\sigma_{\mathrm{tot}}$  exhibits two density regimes. Up to approximately $2-4 \, n_0$, it remains relatively constant at around $100 \, \mathrm{MeV \, fm^{-2}}$; beyond this density threshold, it becomes an increasing function of $n_B$.
The total surface tension is roughly insensitive to variations in $R$ but is affected by changes in $\xi$ and $\eta_V$ starting from $n_B/n_0 \gtrsim 2-3$.
As $\xi$ becomes more negative (see Figs. \ref{fig:surface_tension}a and  \ref{fig:surface_tension}d) and when $\eta_V$ increases (see Figs. \ref{fig:surface_tension}c and \ref{fig:surface_tension}f), an increase in the total surface tension is observed. The effect of $\eta_V$ grows with density and is much more pronounced for Model 1.

In all six panels, we observe that the divergent term has the most significant impact on the total surface tension. The influence of $\sigma_{\mathrm{div}}$ decreases with increasing density, while the effects of all other terms grow with density. Notably,  $\sigma_{\mathrm{det}}$ has the smallest impact and approaches zero beyond approximately $2 n_0$.
The role of $\sigma_{\mathrm{cond}}$ is quite significant because it has a large negative value, especially in the low-density region. As the density increases, this term asymptotically approaches zero and becomes practically negligible beyond $\sim 7 n_0$. The fact that $\sigma_{\mathrm{cond}}$ is always negative can be understood as follows: the quantity $\phi_i^V$ is always negative, as seen in Eq. \eqref{eqq:phi_V}. On the other hand, the quantity $\phi_i^S$ is always positive because $f_{\mathcal{S}, i}(k)$ is always negative (see Eq. \eqref{eqq:phi_S}). Therefore, $\sigma_{\mathrm{cond}}$ is always negative (cf. Eq. \eqref{eqq:sigma_cond}). As the $\phi_i$ tend to zero with increasing density, this causes $\sigma_{\mathrm{cond}}$ to also asymptotically approach zero.
The contribution from $\sigma_\mathrm{free}$ is always positive and increases approximately linearly with density. Notice that $\sigma_\mathrm{free}$ is practically insensitive to variations in $R$ (Figs. \ref{fig:surface_tension}b and \ref{fig:surface_tension}e) and $\eta_V$ (Figs. \ref{fig:surface_tension}c and \ref{fig:surface_tension}f). However, as $\xi$ becomes more negative, $\sigma_{\mathrm{free}}$ increases significantly (Figs. \ref{fig:surface_tension}a and \ref{fig:surface_tension}d).
The contribution from $\sigma_{\mathrm{vec}}$ is always positive and increases approximately linearly with density, becoming  negligible at densities lower than $\sim 3-4 n_0$. The $\sigma_{\mathrm{vec}}$ term is practically insensitive to variations in $R$ (Figs. \ref{fig:surface_tension}b and \ref{fig:surface_tension}e) and $\xi$ (Figs. \ref{fig:surface_tension}a and \ref{fig:surface_tension}d). However, with increasing $\eta_V$, $\sigma_{\mathrm{vec}}$ increases substantially, especially in Model 1.

Finally, in Fig. \ref{fig:curvature_tension} we display the variation of the curvature tension as a function of $n_B/n_0$. The total curvature tension $\gamma_{\mathrm{tot}}$ is shown along with the contributions from $\gamma_{\mathrm{div}}$, $\gamma_{\mathrm{free}}$, $\gamma_{\mathrm{cond}}$, $\gamma_{\mathrm{det}}$, and $\gamma_{\mathrm{vec}}$, as specified in Eq. \eqref{eqq:total_curvature}.
The total curvature tension remains relatively unaffected by changes in $R$ (Figs. \ref{fig:curvature_tension}b and \ref{fig:curvature_tension}e) but it is influenced by variations in $\xi$  (Figs. \ref{fig:curvature_tension}a and \ref{fig:curvature_tension}d) and $\eta_V$ (Figs. \ref{fig:curvature_tension}c and \ref{fig:curvature_tension}f).
Differently from $\sigma_{\mathrm{tot}}$, as $\xi$ becomes more negative  $\gamma_{\mathrm{tot}}$ decreases. However, the dependence with $\eta_V$ follows the same trend as for the surface tension, i.e.  there is an increase in  $\gamma_{\mathrm{tot}}$  as $\eta_V$ increases. The effect of $\eta_V$ grows with density and is much more pronounced for Model 1.

Across all panels, it is evident that the divergent term dominates the total curvature tension. Similar to the behavior observed with surface tension, the influence of the divergent term diminishes as density increases, while the effects of other terms rise with $n_B$. Notably, the determinant term has the least impact and approaches zero beyond approximately $2 n_0$. The condensate term is always negative and its influence is more significant in the low-density region. The contributions from $\gamma_\mathrm{free}$ and $\gamma_{\mathrm{vec}}$ are always positive and increase with density. The influence of $\gamma_{\mathrm{vec}}$ is particularly relevant in Model 1, especially at high densities. All terms are quite insensitive to variations in $R$ and $\eta_V$, except for $\gamma_{\mathrm{vec}}$, which strongly depends on $\eta_V$.

\section{Conclusions}
\label{sec:5}

We have undertaken a comprehensive study of the surface and curvature tensions of three-flavor cold quark matter using the NJL model, incorporating the vector channel into the Lagrangian. Our study ensures both local and global electric charge neutrality, as well as chemical equilibrium under weak interactions. Utilizing the MRE framework to address finite size effects, we examined the influence of particular input parameters, specifically the vector coupling constant, the radius of quark matter droplets, and their charge-per-baryon ratio. Special attention was paid to analyzing how each term in the Lagrangian contributes to the surface and curvature tensions. To the best of our knowledge, this is the first time such an analysis has been performed.

We find that the total surface tension $\sigma_{\mathrm{tot}}$ has two density regimes: it is roughly constant with a value around $100 \, \mathrm{MeV \, fm^{-2}}$  up to $\sim 2-4 \, n_0$ and, above this density, it is an steeply increasing function of $n_B$. The low-density plateau extends while the $s$-quarks are absent in the system. As soon the $s$-quark fraction becomes finite, the total surface tension starts to increase. This is a quite EOS independent behavior associated with the MRE density of states which is quite sensitive to the particle's mass. As emphasized in Ref. \cite{Lugones:2020qll}, the contribution of $s$-quarks to the total surface tension is much larger than the one of $u$ and $d$ quarks. This behavior can be understood if one keeps in mind that for massless particles $f_S=0$  while for $m \rightarrow \infty$, $f_S$ remains finite. This means that light quarks tend to have a smaller contribution to the surface tension.

Our results show that the total surface tension is roughly insensitive to variations in $R$ but is affected by changes in $\xi$ and $\eta_V$ starting from $n_B \gtrsim 2-3 \, n_0$. 
As $\xi$ becomes more negative and as $\eta_V$ increases, an increase in $\sigma_{\mathrm{tot}}$ is observed (see Fig. \ref{fig:surface_tension}). The effect of $\eta_V$ grows with density and is much more pronounced for Model 1.
Regarding the role of each term in the Lagrangian into $\sigma_{\mathrm{tot}}$, we observe that the largest impact comes from the divergent term. This influence decreases with $n_B$, while the effects of all other terms increase with $n_B$. The determinant term, in particular, has a practically negligible effect.
The influence of the condensate terms is quite significant due to their large negative value, especially in the low-density region. As the density increases, this term asymptotically approaches zero and becomes practically negligible beyond $\sim 7 n_0$.
The contributions of $\sigma_\mathrm{free}$ and $\sigma_{\mathrm{vec}}$ are always positive and increase with density, becoming relevant above $\sim 3-4 n_0$.

In contrast to the total surface tension, the curvature tension does not exhibit different density regimes, behaving consistently as a monotonically increasing function of $n_B$. This feature can be understood by examining the $f_C$ contribution to the MRE density of states. As discussed in  \cite{Lugones:2020qll}, for massless particles $f_C=-1 /\left(24 \pi^2\right)$  and for $m \rightarrow \infty$, $f_C$ it is also finite. Consequently, light quarks play a significant role in the curvature contribution,  rendering $\gamma_{\mathrm{tot}}$ less dependent on the emergence of $s$-quarks.

The behavior of the total curvature tension is mostly unaffected by variations in $R$ but is sensitive to changes in $\xi$ and $\eta_V$, as illustrated in Fig. \ref{fig:curvature_tension}. Unlike $\sigma_{\mathrm{tot}}$, $\gamma_{\mathrm{tot}}$ diminishes as $\xi$ becomes more negative. The relationship with $\eta_V$ mirrors that seen with surface tension; specifically, $\gamma_{\mathrm{tot}}$ rises as $\eta_V$ increases. The influence of $\eta_V$ intensifies with density.

Regarding the influence of each term in the Lagrangian on the total curvature tension, the divergent term clearly dominates $\gamma_{\mathrm{tot}}$. Similar to the behavior observed with surface tension, the influence of $\gamma_{\mathrm{div}}$ diminishes as density increases, while the contributions from other terms rise with $n_B$. Notably, $\gamma_{\mathrm{det}}$ is negligible. The condensate term is always negative and becomes more significant in the low-density region. The contributions from $\gamma_\mathrm{free}$ and $\gamma_{\mathrm{vec}}$ are consistently positive and increase with density. The role of $\gamma_{\mathrm{vec}}$ is is quite relevant in Model 1, especially at high densities. All contributions are relatively insensitive to variations in $R$ and $\eta_V$, except for $\gamma_{\mathrm{vec}}$, which strongly depends on $\eta_V$.

When comparing Model 1 and Model 2, it is noteworthy that the forms of the vector terms differ significantly. In Model 1, the vector term incorporates the square of the sum of the $n_i$, resulting in six quadratic terms. Conversely, in Model 2, the vector term is directly the sum of the $n_i^2$, yielding only three quadratic terms. This distinction directly influences the growth of $\sigma_{\mathrm{vec}}$ and $\gamma_{\mathrm{vec}}$ with density. In fact, in Model 1, these quantities increase much more with density than in Model 2. Therefore, to achieve comparable increases in both models for a given density, the value of $\eta_V$ used in Model 2 should be significantly higher than in Model 1 to compensate for the missing terms.

Another interesting aspect that arises from our results relates to the behavior of $\sigma_{\mathrm{tot}}$ in the low-density regime. Indeed, in the density range up to approximately $3-4 n_0$, $\sigma_\mathrm{tot}$ shows a plateau with a value of about $100 \, \mathrm{MeV \, fm^{-2}}$ which is quite insensitive to variations in the parameters $\xi$, $R$, and $\eta_V$. This result is quite interesting because it shows that the hypothesis of constant surface tension, used in several works in the literature to describe the quark-hadron mixed phase, could be a reasonable approximation in the low-density regime.

Finally, it is worth comparing the results from the NJL model with those obtained in previous works using the MIT bag model. Since the MIT bag model does not incorporate dynamic masses, this comparison is particularly meaningful at the highest possible densities, which is precisely where the dynamic masses in the NJL model approach their current masses.
Comparing the NJL Lagrangian of this work with that of the vector MIT bag model from Ref. \cite{Lugones:2021tee}, we see that both the vector and the ``free" terms are present in both Lagrangians. Indeed, the vector term used in Ref. \cite{Lugones:2021tee} has the same functional form as in Model 1 of the present work. On the other hand, both the determinant term and the condensate term in the NJL model are practically negligible at high enough densities. Therefore, the main difference between both models at high densities arises from the divergent term of the NJL model.
If we compare the value of $\sigma$ from the MIT bag model, as given in \cite{Lugones:2021tee}, with $\sigma_\mathrm{free} + \sigma_\mathrm{vec,1}$ from the NJL model, we find that the values are quite similar, and the same applies to $\gamma$. In fact, at high enough densities, the major difference in the total surface and curvature tensions between these models stems from the dominant contribution of the divergent term in the NJL model.

\section{Acknowledgements}
This work has been partially funded by CONICET (Argentina) under Grant No. PIP 2022-2024 GI - 11220210100150CO, by ANPCyT (Argentina) under Grants No. PICT17-03-0571 and PICT20-01847, and by the National University of La Plata (Argentina), Project No. X824.
G.L. acknowledges the financial support from the Brazilian agencies CNPq (grant 316844/2021-7) and FAPESP (grant 2022/02341-9).

\bibliography{references}

\end{document}